\documentclass[%
 reprint,
 superscriptaddress,
 amsmath,amssymb,
aps,
]{revtex4-1}

\usepackage{graphicx}
\usepackage{dcolumn}
\usepackage{bm}

\usepackage{mystyle}
\usepackage{libertine}

\begin{document}

\preprint{APS/123-QED}

\title{Probing nonlocal superconducting fluctuations with covariance noise magnetometry}

\author{Gustav Romare}
\affiliation{Department of Physics, University of Wisconsin-Madison, Madison, Wisconsin 53706, USA}

\author{Ilya Esterlis}
\affiliation{Department of Physics, University of Wisconsin-Madison, Madison, Wisconsin 53706, USA}

\author{Shimon Kolkowitz}
\affiliation{University of California at Berkeley, Department of Physics, Berkeley CA 94720, USA}

\author{Alex Levchenko}
\affiliation{Department of Physics, University of Wisconsin-Madison, Madison, Wisconsin 53706, USA}

\date{\today}

\begin{abstract}
The nonlocal superconducting fluctuation corrections to the conductivity tensor $\sigma_{ij}(\bfq,\omega)$ are calculated within the time-dependent Ginzburg-Landau framework, and their observable consequences for quantum noise magnetometry are worked out. For a single nitrogen-vacancy (NV) sensor we obtain the relaxation rate $1/T_1$ as a function of temperature, sample-sensor distance, and probe frequency, identifying the scales at which the nonlocality and the dynamics of the pair fluctuations cut off the critical enhancement near $T_c$. For two-sensor covariance magnetometry we show that the two-point field correlator develops additional spatial structure whose range directly measures the fluctuation correlation length $\xi(T)$. We further analyze two channels that accompany the paraconductivity: the Maki-Thompson correction to the spin susceptibility, and the fluctuation diamagnetism. Finally, we solve exactly, to all orders in a dc electric field and at all wave vectors, for the nonequilibrium current noise of the fluctuating film: the noise decouples from the nonlinear paraconductivity, violating the fluctuation-dissipation theorem by universal factors at criticality and acquiring a bias-induced spatial anisotropy directly measurable by covariance magnetometry. The results are connected to a recent experiment measuring current noise near a thin film of BSCCO.
\end{abstract}

\maketitle

\section{Introduction}

Near the superconducting transition, reduced dimensionality and short coherence lengths enhance pair fluctuation effects, which manifest in both thermodynamic and dynamical properties \cite{larkin2005,varlamov2018}. Fluctuation effects are well-documented in a variety of systems, including amorphous \cite{glover1971,pourret2006} and crystalline \cite{kajimura1971,hsu1992} thin films of conventional superconductors \cite{skocpol1975}, high-$T_c$ cuprates \cite{vidal1988,rullier2011,cimberle1997,hopfengartner1991,mandal1990,duan1991}, and superconductors based on two-dimensional van der Waals materials \cite{song2024}.

Fluctuation corrections to the conductivity in particular arise through the Aslamazov-Larkin (AL) paraconductivity \cite{aslamazov1968}, Maki-Thompson (MT) interference effects \cite{maki1968,thompson1970}, and density of states suppression \cite{abrahams1970}. In general, the fluctuation conductivity---which encodes the low-energy dynamics of superconductors near $T_c$---is a nonlocal, dynamical response function, $\sigma_{ij}(\bfq, \omega)$, that depends on both wave vector $\bfq$ and frequency $\omega$. Conventional dc transport measurements probe only its long-wavelength, low-frequency limit, $\bfq \to 0$ and $\omega \to 0$, and consequently access only a small part of the information contained in the full fluctuation response. That information is considerable: the momentum dependence of $\sigma_{ij}(\bfq,\omega)$ encodes the fluctuation correlation length $\xi(T)$ and its critical divergence, while the frequency dependence encodes the order-parameter relaxation time $\tau_{\rm GL}$---the critical slowing down near the transition. A probe with access to finite $(\bfq,\omega)$ can therefore measure static and dynamical critical properties, distinguish Gaussian from vortex-dominated (Berezinskii-Kosterlitz-Thouless) fluctuation regimes through the functional form of $\xi(T)$, separate intrinsic critical correlations from static, disorder-induced inhomogeneity, and---once driven out of equilibrium---expose physics with no counterpart in linear transport, such as the breakdown of the fluctuation-dissipation theorem (FDT) in the pair-fluctuation channel. Developing this program quantitatively is the purpose of the present paper.

\begin{figure}[t!]
    \centering
    \includegraphics[width=\linewidth]{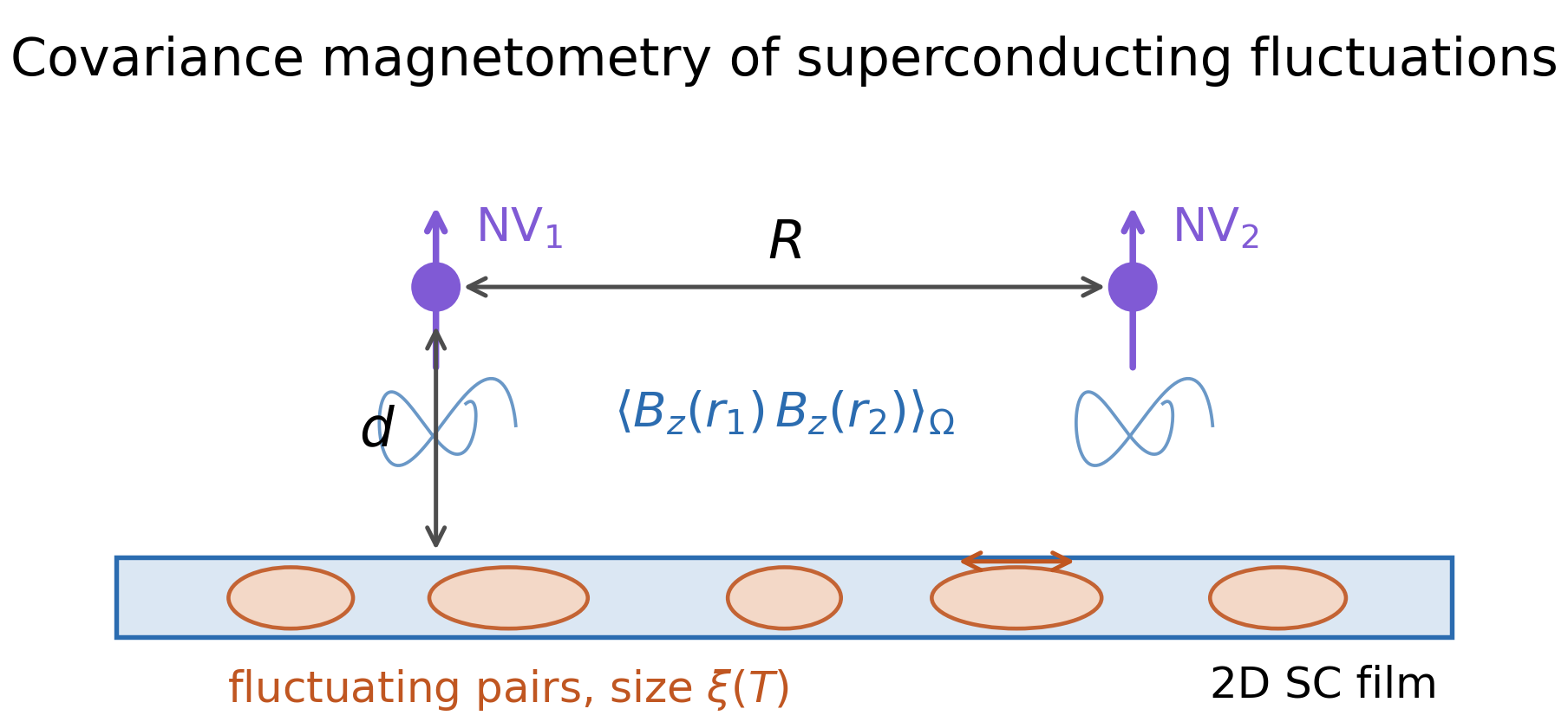}
    \caption{Schematic of the setup. Two NV sensors at height $d$ above a two-dimensional superconductor, separated by an in-plane distance $R$, detect the magnetic noise generated by fluctuating Cooper pairs of size $\xi(T)$. Single-sensor relaxometry measures the local field autocorrelation; covariance magnetometry measures the two-point correlator $\langle B_z(\mathbf r_1)B_z(\mathbf r_2)\rangle_\omega$, which resolves the spatial structure of the fluctuations.}
    \label{fig:setup}
\end{figure}

Recent measurements of current noise near thin films of BSCCO have revealed signatures of pronounced superconducting fluctuations around $T_c$ \cite{liu2025}; related experiments have also been recently carried out on Nb \cite{li2026}. These experiments measured the relaxation rate $1/T_1$ of nitrogen-vacancy (NV) centers---point-like defects in diamond that act as single-spin (or qubit) magnetometers---placed near the sample surface as a function of temperature $T$. Magnetic field sensing with NV centers has been used in recent years to study a variety of static and dynamical phenomena in condensed matter systems  \cite{casola2018probing,rovny2024,thiel2019probing,bhattacharyya2024,ku2020,du2017,andersen2019}. In particular, measurements of NV relaxation rates have been proposed as a noninvasive probe of material properties, such as nonlocal conductivity, that are challenging to extract via conventional methods \cite{agarwal2017,rustagi2020,machado2023}, with example systems including superconductors \cite{chatterjee2022,dolgirev2022,curtis2024}, magnetic insulators \cite{chatterjee2019}, one-dimensional quantum liquids \cite{rodriguez2018}, and electron solids \cite{dolgirev2024}. A schematic of the experimental setup is shown in Fig.~\ref{fig:setup}.

Here, we calculate the AL contribution to $\sigma_{ij}(\bfq,\omega)$ and the corresponding magnetic field noise generated near the sample, using the time-dependent Ginzburg-Landau (TDGL) formalism \cite{schmid1966,larkin2005}. Closed-form results for the nonlocal fluctuation electromagnetic response were in fact obtained within TDGL already in the early 1990s, see Ref. \cite{barash1993} for bulk superconductors and Ref. \cite{galaktionov1995} for thin films \cite{galaktionov1995}; we make contact with those results below and extend them to the noise observables of interest here. We reproduce the theoretical results presented in Ref.~\cite{liu2025} for the dependence of $1/T_1$ on temperature, and extend them to include the variation of the relaxation rate with qubit-sample distance and qubit frequency. We also analyze the contribution to magnetic noise from spin (as opposed to current) fluctuations. We then show that the recently developed covariance magnetometry technique \cite{rovny2022,le2025,rovny2025,cheng2025,huxter2025,cambria2025,hosseinabadi2026} provides a particularly promising route for measuring nonlocal pair fluctuation effects. Going beyond the paraconductivity, we also quantify the fluctuation diamagnetism and argue that it is measurable with current NV magnetometers. Finally, we solve exactly for the \textit{nonequilibrium} current noise of the fluctuating film under a dc bias, obtaining the universal violation of the FDT in the pair-fluctuation channel.

The paper is organized as follows. Section~\ref{sec:nonlocal} reviews the TDGL calculation of the nonlocal AL conductivity. Section~\ref{sec:singleNV} develops the single-NV relaxometry, including the dependence on distance, frequency, and the spin (MT) channel. Section~\ref{sec:covariance} treats two-sensor covariance magnetometry. Section~\ref{sec:diamagnetism} analyzes fluctuation diamagnetism, and Sec.~\ref{sec:noneq} the nonequilibrium noise and FDT breakdown. Section~\ref{sec:summary} summarizes the main results, and Sec.~\ref{sec:discussion} concludes with a discussion. Further technical details are relegated to Appendices \ref{app:noise}--\ref{app:dia}.

\section{Nonlocal fluctuation conductivity}
\label{sec:nonlocal}

In this section, we briefly review the TDGL approach to AL paraconductivity and summarize the connection between the nonlocal conductivity and the relevant quantities measured in NV experiments.

The AL contribution can be captured by TDGL theory, which gives the order parameter dynamics as
\begin{equation}\label{TDGL-Linearized}
\left[\gamma\frac{\partial}{\partial
t}-\frac{\nabla^{2}}{2m}+a\right]\Psi(\mathbf{r},t)=\zeta(\mathbf{r},t).
\end{equation}
Here $\gamma$ determines the order-parameter relaxation rate, $a(T) = \alpha(T/T_c-1)$ and $\zeta(\mathbf r, t)$ is a Langevin force with white noise correlations, $\langle \zeta(\mathbf r, t)\zeta^*(\mathbf r', t')\rangle = 2\gamma T \delta(\mathbf r - \mathbf r')\delta(t-t')$. Here and below, $e$ and $m$ denote the charge and mass of a Cooper pair ($e \to 2e_0$ and $m \to 2m_e$ in terms of the electron values) and we have set $\hbar = 1$, $k_B =1$. Eq.~\eqref{TDGL-Linearized} can be solved using the Green's function method
\begin{equation}
\Psi(\mathbf{r},t)=\int\mathcal{G}(\mathbf{r},t;\mathbf{r}',t')\zeta(\mathbf{r}',t')
\mathrm{d}^{d}\mathbf{r}'\mathrm{d}t',
\end{equation}
with $\mathcal G$ given in Fourier representation by
\begin{equation}\label{GreenFunctionFourier}
\mathcal{G}(\mathbf{q},\omega)=\frac{1}{\varepsilon_{\mathbf{q}}-i\gamma\omega},
\quad\varepsilon_{\mathbf{q}}=\frac{\mathbf{q}^{2}}{2m}+a.
\end{equation}
With this we can also construct the fluctuation propagator,
\begin{equation}\label{FluctuatingPropagator-Int}
\Pi(\mathbf{x}_{1};\mathbf{x}_{2})=2\gamma
T\int\mathcal{G}(\mathbf{r}_{1},t_{1};\mathbf{r}'_{1},t'_{1})
\mathcal{G}^{*}(\mathbf{r}_{2},t_{2};\mathbf{r}'_{1},t'_{1})
\mathrm{d}^{d}\mathbf{r}'_{1}\mathrm{d}t'_{1}
\end{equation}
which has the Fourier representation
\begin{equation}\label{FluctuatingPropagator-FourierResult}
\Pi(\mathbf{p},\varepsilon)=\frac{2\gamma
T}{\varepsilon^{2}_{\mathbf{p}}+\gamma^{2}\varepsilon^{2}}.
\end{equation}
We are now in a position to calculate the conductivity tensor. The fluctuation dissipation theorem relates the current fluctuation $C^{JJ}_{ij} = 1/2\langle\{J_i,J_j\}\rangle=\langle J_iJ_j\rangle$ to the retarded response $\chi^{JJ}_{ij}$ by
\begin{equation}\label{current-fdt}
    C^{JJ}_{ij}(\mathbf q,\omega) = \coth\left(\frac{\omega}{2T}\right)\text{Im}\left\{\chi^{JJ}_{ij}(\mathbf q.\omega)\right\}
\end{equation}
The retarded response can be related to the conductivity tensor as $ \chi^{JJ}_{ij}=i\omega \sigma_{ij}$ and we find
\begin{equation}
    \text{Re}\, \sigma_{ij}(\mathbf q, \omega) = \frac{1}{2T}C^{JJ}_{ij}(\mathbf q, \omega),
\end{equation}
where we expanded the hyperbolic function in $\omega/T \ll 1$. The current $J_i$ entering the current correlation function is given by the usual GL expression,
\begin{equation}\label{Current-Def}
J_{i}(\mathbf{r},t)=\frac{e}{2mi}[\Psi^{*}\nabla_{i}\Psi-\Psi\nabla_{i}\Psi^{*}].
\end{equation}
In Fourier space, this becomes
\begin{equation}
J_{i}(q)=\frac{e}{m}\int
p_{i}\Psi(p+q/2)
\Psi^{*}(p-q/2)
\frac{\mathrm{d}^{d+1}p}{(2\pi)^{d+1}},
\end{equation}
where we used 4-vector notation $q\equiv (\mathbf q, \Omega)$ and $p\equiv (\mathbf p, \varepsilon)$. The conductivity tensor can now be expressed using the fluctuation propagator as
\begin{equation}\label{ConductivityTensor}
\text{Re}\,\sigma_{ij}(k)=\frac{e^{2}}{2m^{2}T}\int\
p_{i}p_{j}\Pi(p+k/2)
\Pi(p-k/2)\
\frac{\mathrm{d}^{d+1}p}{(2\pi)^{d+1}}.
\end{equation}
The imaginary part of the conductivity can be restored by Kramers-Kronig. The integrals can be performed for general momentum and frequency (see Appendix~\ref{app:AL}); for brevity, below we will quote the $d=2$ static results. Assuming an isotropic system, the conductivity can be decomposed into its transverse and longitudinal parts
\begin{equation*}
    \sigma_{ij}(\kappa) = \sigma^{AL}\left[(\delta_{ij}-\hat\kappa_i\hat\kappa_j)F_T^{(2D)}(\kappa) + \hat\kappa_i\hat\kappa_jF_L^{(2D)}(\kappa)\right],
\end{equation*}
with the dimensionless functions,
\begin{align}
    F_L^{(2D)}(\kappa)&=\frac{\ln(1+\kappa^2)}{\kappa^2}\;,\\
    F_T^{(2D)}(\kappa)&=\frac{2}{\kappa\sqrt{1+\kappa^2}}\,\mathrm{arcsinh}\,\kappa-\frac{\ln(1+\kappa^2)}{\kappa^2}. \label{eq:ft}
\end{align}
Here $\kappa = \mathbf k \xi/2$ is a dimensionless momentum vector with $\xi^2(T) = 1/2ma$ being the superconducting correlation length and
\begin{equation}\label{eq:sigmaAL}
    \sigma^{\text{AL}} =\frac{ \gamma k_BTe^{2}m\xi^{2}}{4\pi}.
\end{equation}
With the microscopic value of the TDGL relaxation constant, $\gamma = \pi\alpha/8k_BT_c$ \cite{larkin2005}, Eq.~\eqref{eq:sigmaAL} reduces to the universal Aslamazov-Larkin sheet conductivity $\sigma^{\rm AL} = e_0^2/16\hbar\epsilon$ (restoring $\hbar$), with $e_0$ the electron charge and $\epsilon=(T-T_c)/T_c$. The static transverse kernel Eq.~\eqref{eq:ft} coincides with the thin-film fluctuation response derived in Ref. \cite{galaktionov1995}; the correspondence extends to finite frequency and to the reactive (diamagnetic) part of the kernel, as discussed in Sec.~\ref{sec:diamagnetism} and Appendix~\ref{app:dia}. Below we will make use of the limiting behaviors of function $F_T$:
    \be
    F_T(\kappa) \to \begin{cases} 1-5\kappa^2/6, & \kappa \ll 1, \\ \ln(4)/\kappa^2, & \kappa \gg 1. \end{cases}
    \ee

The relevant physical observable for both qubit noise spectroscopy and covariance magnetometry based on NV centers is the magnetic noise tensor
    \begin{align}
        \mathcal N_{\alpha\beta}(\bfr, \omega) = \frac 12 \int_{-\infty}^\infty \dd t ~ e^{i\omega t} \langle \{ B_\alpha(\bfr,t), B_\beta(0,0)\}\rangle.
    \end{align}
The noisy magnetic field $B_\alpha(\mathbf r,t)$ is related to fluctuating currents in the nearby 2D system by the Biot-Savart law. Combining this relation with the fluctuation-dissipation theorem yields an expression for the noise  tensor in terms of the transverse conductivity \footnote{Strictly speaking, this formula also assumes $\Omega d \ll c$, where $c$ is the speed of light in vacuum. For $\Omega \sim \text{GHz}$ and $d \sim $tens of nm---the typical operating regime of single NV centers---this inequality is well-satisfied.} (see Appendix \ref{app:noise}) . We focus on the component
\begin{align}
    \mathcal N_{zz}(\mathbf R, z,
    \omega) &= \frac{\mu_0^2k_BT}{4\pi}\int_0^\infty\dd q\, qJ_0(Rq)e^{-2qz}\sigma'_T(q,\omega). \label{eq:Nzz_Rzw}
\end{align}
Here $\bfr = (\bfR,z)$, with $\bfR$ the 2D position vector in the plane of the material and $z$ the distance above the plane; $\sigma'_T$ denotes the real part of the transverse conductivity.

\section{Local noise}
\label{sec:singleNV}

\begin{figure}[h]
    \centering
    \includegraphics[width=\linewidth]{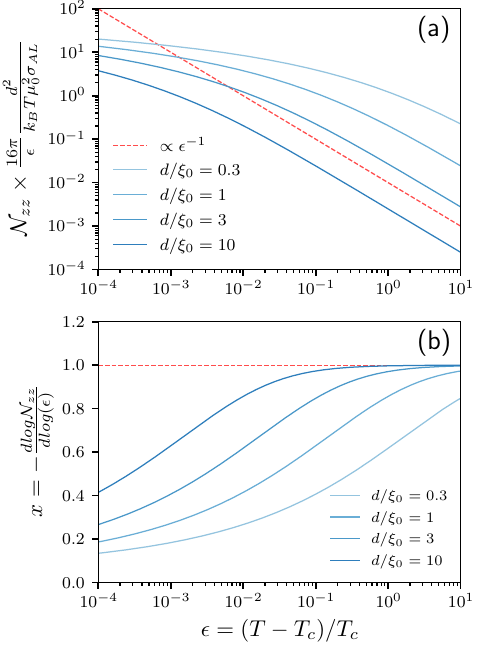}
    \caption{(a) Transverse noise as a function of reduced temperature $\epsilon = (T-T_c)/T_c$ for different values of NV distance $d/\xi_0$. The dashed line corresponds to $\epsilon^{-1}$ divergence. (b) Effective exponent $x = -d\ln(\mathcal N_{zz})/d\ln \epsilon$ of the divergence near $T_c$.}
    \label{fig:AL_T1}
\end{figure}

\begin{figure}[h]
    \centering
    \includegraphics[width=\linewidth]{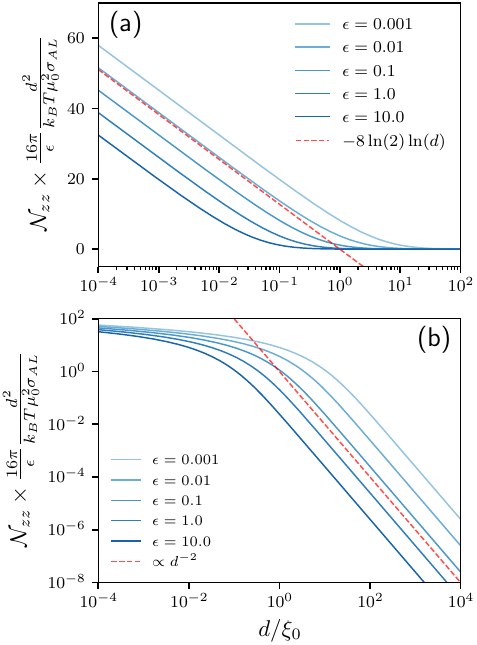}
    \caption{(a) Scaling behavior for $d/\xi_0\ll 1$. The graph shows the local noise $\mathcal N_{zz}$ as a function of NV separation in units of the coherence length $d/\xi_0$ for different values of the reduced temperature. The red dashed line is the logarithmic limiting behavior in Eq.~\eqref{eq:T1limits}. (b) Scaling relation for $d/\xi_0\gg 1$. The red dashed line is the $\xi^2(T)/4d^2$ limiting behavior in Eq.~\eqref{eq:T1limits}.}
    \label{fig:T1_distance_scaling}
\end{figure}

\begin{figure}
    \centering
    \includegraphics[width=\linewidth]{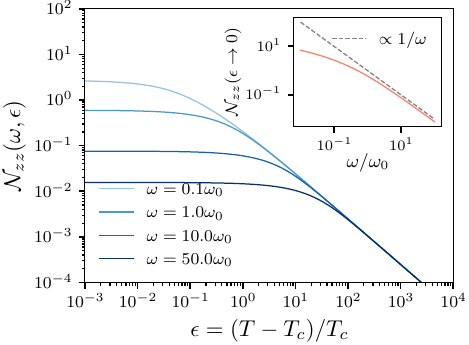}
    \caption{Noise for various finite probe frequencies $\omega$. The noise saturates to a finite value below a reduced temperature $\epsilon_\omega \sim \omega/\omega_0$, where $\omega_0 = 1/m\gamma\xi_0^2$. Inset: The frequency dependence of the noise at $T_c$. For large $\omega/\omega_0$, the noise decays $\propto 1/\omega$.}
    \label{fig:noise_frequency_scaling}
\end{figure}

We start with the case of noise detection at a single point above the plane, which corresponds to setting $\mathbf R=0$ in \eqref{eq:Nzz_Rzw}. The local noise $\mathcal N_{zz}(z, \omega) \equiv \mathcal N_{zz}(\bfR=0,z, \omega)$ is directly related to the relaxation rate $1/T_1$ of a NV center placed a distance $z=d$ above the material with level-splitting $\omega$ \cite{langsjoen2012,agarwal2017}; for details see Appendix \ref{app:noise}. We have
\begin{equation}
    \mathcal N_{zz}(z=d,
    \omega) = \frac{\mu_0^2k_BT}{4\pi}\int_0^\infty\dd q\, qe^{-2qd}\sigma'_T(q,\omega). 
\label{eq:T1kernel}
\end{equation}
Here and below we restore $k_B$. 
For a rough estimate of the noise, note that the integration kernel is peaked at a wave vector $q \sim 1/2d$, so that
\begin{equation}
\mathcal N_{zz}(d,\omega) \sim \frac{\mu_0^{2}k_BT}{d^{2}}\,\sigma'_T\!\Big(q_\star=\frac{1}{2d},\,\Omega\Big).
\label{eq:T1single}
\end{equation}
Local noise measurements thus probe the nonlocal conductivity at a single wave vector 
selected by the sample-to-detector distance.

We now consider the AL correction to $\mathcal N_{zz}(d,\omega)$. Here we observe that, in an NV experiment, the frequency $\omega$ is typically a few GHz, so that the relevant dimensionless combination $\omega\tau_{\rm GL} \ll 1$ over most of the accessible temperature range and we may set the frequency to zero in $\sigma'_T$ (see below for a discussion of the frequency dependence). Inserting the zero-frequency result \eqref{eq:ft} into \eqref{eq:T1kernel} we obtain the AL fluctuation correction
\begin{equation}
    \mathcal N^{\rm AL}_{zz}(d) = \frac{\mu_0^2}{16 \pi} 
    \frac{k_BT \sigma^{\rm AL}}{ d^2}\int_0^\infty \dd x\, xe^{-x}F_T\left(\frac{\xi_0x}{4d\sqrt\epsilon}\right)
    \label{eq:AL_T1_contribution}
\end{equation}
where $\mathcal N_{zz}^{\rm AL}(d) \equiv \mathcal N_{zz}^{\rm AL}(d,\omega=0)$ and $\xi_0 = 1/\sqrt{2m\alpha}$, so that $\xi(T) = \xi_0/\sqrt\epsilon$.

For reference, we also record the relaxation rate deep in the metallic normal state,
$\mathcal N^0_{zz}(d,\omega)$, where the qubit relaxation is due to evanescent Johnson noise \cite{langsjoen2012,kolkowitz2015}
    \begin{equation}
    \mathcal N^0_{zz}(d,\omega = 0) = \frac{\mu_0^2}{16 \pi}
    \frac{k_BT \sigma_0}{d^2}.
        \label{eq:T1metal}
    \end{equation}
Here $\sigma_0$ is the ordinary dc sheet conductivity of the metal; this follows from Eq.~\eqref{eq:T1kernel} with $\sigma'_T = \sigma_0$ and $\int_0^\infty q e^{-2qd}\dd q = 1/4d^2$.

We begin by investigating the expression Eq.~\eqref{eq:AL_T1_contribution} for the fluctuation contribution to the local noise in different limits. The argument of $F_T$ in the integrand contains the mean-field correlation length $\xi(T) =\xi_0/\sqrt\epsilon$, setting the natural length scale for the ``close" detector $d < \xi(T)$ and the ``distant" detector $d > \xi(T)$ regimes.
In the extreme limits, we find from Eq.~\eqref{eq:AL_T1_contribution}
\begin{equation}
    \mathcal N^{\rm AL}_{zz}(d) \sim \sigma^{\rm AL}\times \begin{cases} \dfrac{1}{4d^2}\left[1 - \dfrac{5\,\xi^2(T)}{16\, d^2}\right], & d\gg \xi(T), \\[2ex]
    \dfrac{8\ln 2}{\xi^2(T)}\,\ln\!\left[\dfrac{\xi(T)}{c_1\,d}\right], & d \ll \xi(T),
    \end{cases}
    \label{eq:T1limits}
\end{equation}
where the constant $c_1$ under the logarithm is determined numerically as $c_1 \approx 9.1$ 
Since $\sigma^{\rm AL}\propto\xi(T)^2 \sim 1/\epsilon $, we see that, for a fixed detector distance $d$, the local noise---and hence the single-NV relaxation rate---will grow as $1/(T-T_c)$ as the system is cooled from the normal state; this behavior was observed in the $1/T_1$ measurements of Ref.~\cite{liu2025}. This growth with decreasing $T$ will continue until $\xi(T) > d$, when the noise will cross over to the much slower logarithmic growth (note that $\sigma^{\rm AL}/\xi^2(T)$ is temperature independent); this regime is sensitive to the $q$-dependence of $\sigma_T'$. The offset constant in Eq.~\eqref{eq:T1limits} shows that the logarithmic regime is fully developed only once $\xi(T) \gtrsim 10\, d$.

In Fig.~\ref{fig:AL_T1}a we plot our numerical results for $\mathcal N_{zz}^{\rm AL}(d)$ obtained by numerical integration of Eq.~\eqref{eq:AL_T1_contribution}, where the limiting behaviors \eqref{eq:T1limits} are also verified. In Fig.~\ref{fig:AL_T1}b we also show the ``apparent exponent'' in the growth of the noise as $T \to T_c$, $x = -d\log(\mathcal N_{zz}^{\rm AL})/d\log\epsilon$. From the figure it is clearly seen that the $1/(T-T_c)$ scaling of the noise persists over a broader temperature range the further the detector is from the sample.

The probe frequency $\omega$ also defines a certain length scale $\xi_\omega \sim \sqrt{k_B T_c/\hbar \omega}\times \xi_0$ that serves to cut off the divergence of the noise as $T \to T_c$; a similar observation regarding the ac conductivity was made in \cite{dorsey1991}. The length scale $\xi_\omega$ corresponds roughly to the distance over which a fluctuation diffuses over the time scale $1/\omega$ set by the probe frequency. This translates to a reduced temperature $\epsilon_\omega \sim \hbar\omega/k_B T_c$ below which the noise saturates to a finite value controlled by $\omega$. The saturation criterion can equivalently be stated as  $\omega\tau_{\rm GL} \sim 1$, with $\tau_{\rm GL} = \pi\hbar/8k_B(T-T_c)$ the GL relaxation time. The evolution of the noise as $T \to T_c$ for non-zero frequencies is shown in Fig.~\ref{fig:noise_frequency_scaling}. Close to $T_c$, the noise decreases with frequency as $\mathrm{Re}\, F_T \sim 1/\omega$ at large $\omega$ (see inset of Fig.~\ref{fig:noise_frequency_scaling} and Appendix~\ref{app:AL},  Fig.~\ref{fig:conddyn}). Using a probe frequency $\omega \approx 3 ~ {\rm GHz}$ and $T_c \approx 90$ K appropriate for the BSCCO NV experiment \cite{liu2025}, we estimate  a reduced temperature $\epsilon_\omega \sim 10^{-3}$ below which noise should saturate. Since the NV splitting is field-tunable over $\sim1$--$5$~GHz, frequency-resolved relaxometry directly measures the critical slowing down $\tau_{\rm GL}(\epsilon)$, a quantity inaccessible to dc transport.

\subsection{Spin channel: Maki-Thompson contribution}
\label{sec:spin}

Except in very clean metals, the MT corrections to the fluctuation \textit{conductivity} are much smaller than the AL corrections and we have thus ignored them. However, NV relaxation is also influenced by magnetic noise from spin fluctuations in the material \cite{agarwal2017,chatterjee2019}, which are in turn related to the imaginary part of the spin susceptibility $\chi''$, and in this channel the fluctuation physics enters through the anomalous MT process---the pairing of an electron with its time-reversed partner---analyzed for the NMR relaxation rate by Randeria and Varlamov \cite{randeria1994}. The MT correction to the dissipative spin response is built from the product of the fluctuation propagator and the Cooperon and, in two dimensions, takes the form
    \begin{equation}
        \frac{\delta\chi''_\text{MT}(\bfq,\omega)}{\omega}\bigg|_{\omega\to0} = \frac{\chi''_0(\omega)}{\omega}\, \beta_{\rm MT}\, \frac{\ln(\epsilon/\gamma_\phi)}{\epsilon - \gamma_\phi}\, \mathcal K(q),
        \label{eq:MTchi}
    \end{equation}
where $\chi''_0/\omega = \pi\nu_0^2$ is the normal-state (Korringa) value, $\gamma_\phi = \pi\hbar/8k_BT\tau_\phi$ is the pair-breaking parameter set by the phase-relaxation time $\tau_\phi$, $\beta_{\rm MT} = \mathcal O(Gi)$ is a positive constant of order the Ginzburg number, and $\mathcal K(q)$ is a form factor with $\mathcal K(0) = 1$ that decays for $q \gtrsim \min(\xi^{-1}, L_\phi^{-1})$, with $L_\phi = \xi_0/\sqrt{\gamma_\phi}$ the dephasing length. The singular factor follows from the elementary pair-momentum integral $\int\! \frac{\dd^2Q}{(2\pi)^2}[(\epsilon + \xi^2Q^2)(\gamma_\phi + \xi^2Q^2)]^{-1} = \ln(\epsilon/\gamma_\phi)/[4\pi\xi^2(\epsilon-\gamma_\phi)]$. Considering, for simplicity, the regime where the detector height $d > \xi$, the contribution to the noise from spin fluctuations is $\mathcal N_s \sim \mu_0^2\mu_B^2 T/d^4 \times [\delta \chi_{\rm MT}''(0,\omega)/\omega]_{\omega \to 0}$.

Two features are noteworthy. First, the MT enhancement grows as $\sim(1/\epsilon)\ln(\epsilon/\gamma_\phi)$ for $\epsilon \gg \gamma_\phi$ but \textit{saturates} at $\epsilon \sim \gamma_\phi$: pair breaking cuts off the divergence, in analogy with the distance and frequency cutoffs of the orbital channel. Second, for a sign-changing ($d$-wave) order parameter, as in BSCCO, the anomalous MT term is strongly suppressed, and the accompanying negative density-of-states correction, $\delta(1/T_1T)_{\rm DOS} \propto -\ln(1/\epsilon)$, can dominate the spin channel \cite{randeria1994}---so that a near-$T_c$ \textit{peak} in the NV relaxation is a fingerprint of the orbital (paraconductivity) channel.
 
For the two-sensor covariance introduced in the next section, the same structure implies a spin-channel covariance
\begin{equation}
    \mathcal N_s(R) \propto \frac{K_0(R/\xi) - K_0(R/L_\phi)}{L_\phi^{-2} - \xi^{-2}},
    \label{eq:Cspin}
\end{equation}
a difference of modified Bessel functions whose spatial range is set by the \textit{longer} of $\xi(T)$ and $L_\phi$. In the regime $\epsilon > \gamma_\phi$, where the MT enhancement operates, one has $\xi(T) < L_\phi$, so the spin covariance decays on the dephasing length---which is only weakly temperature dependent---in sharp contrast to the orbital channel, whose range tracks the strongly temperature-dependent $\xi(T)$. Covariance magnetometry can thus, in principle, disentangle the two channels by their range and its temperature dependence, and thereby extract the electronic dephasing length.

\section{Two-point correlations and connection to covariance magnetometry}
\label{sec:covariance}

In NV covariance magnetometry \cite{rovny2022}, two NV centers are read out simultaneously, shot by shot, and the observable is the covariance of the two photon records, normalized as the Pearson coefficient $r = \mathrm{Cov}(S_1,S_2)/\sigma_1\sigma_2$. Because the local photon shot noise and single-NV spin-projection noise are uncorrelated between the sensors, they cancel from the covariance, which isolates the \textit{shared} magnetic signal produced by the sample. For phase-accumulation (Ramsey- or echo-type) protocols with filter function $W(t)$, the covariance of the accumulated phases measures $\gamma_{\rm NV}^2\int \frac{\dd\omega}{2\pi}|\tilde W(\omega)|^2\, \mathcal N_{zz}(R,d,\omega)$, with $\gamma_{\rm NV} = g\mu_B/\hbar$; for relaxometry-mode covariance, the correlated part of the two decay rates measures the cross-spectral density at the NV splitting \cite{le2025,hosseinabadi2026,rovny2025}. In either mode the measured correlations are governed by the two-point magnetic correlation function in Eq.~\eqref{eq:Nzz_Rzw},  with $r(R) = \mathcal N_{zz}(R)/[\mathcal N_{zz}(0) + N_{\rm loc}]$, with $N_{\rm loc}$ the local (uncorrelated) noise floor.
Notice tha the Bessel factor $J_0(qR)$ in Eq.~\eqref{eq:Nzz_Rzw} oscillates on the scale $q\sim 1/R$, so that varying the sensor separation scans the nonlocal conductivity in wave-vector space. We show below that, where the single-sensor rate saturates once $\xi(T)$ exceeds $d$ (see Eq.~\eqref{eq:T1limits}), the covariance resolves the very structure responsible for that saturation

\textit{(i) Metallic regime.} Deep in the metallic phase (or, near $T_c$, whenever $\xi(T)\ll d$), the conductivity is local at the probed momenta, $\sigma'_T \to \sigma_0$, and the transform \eqref{eq:Nzz_Rzw} is elementary:
\begin{equation}
    \frac{\mathcal N_{zz}(R)}{\mathcal N_{zz}(0)} = \left[1 + \left(\frac{R}{2d}\right)^2\right]^{-3/2}.
    \label{eq:czzlocal}
\end{equation}
The covariance is flat for $R \lesssim 2d$ and falls as $1/R^3$ for $R \gg 2d$---a purely geometric profile whose range is set by the standoff distance alone and is strictly temperature independent; only the overall amplitude, $\propto k_BT\sigma_0$ (Johnson noise), varies. (We ignore the ballistic regime where the mean free path $\ell > d$ \cite{kolkowitz2015}, which is irrelevant for the materials of interest.) A  temperature-dependent change in the \textit{shape} of $r(R)$ therefore signals a growing correlation length.

\begin{figure}[t]
    \centering
    \includegraphics[width=\linewidth]{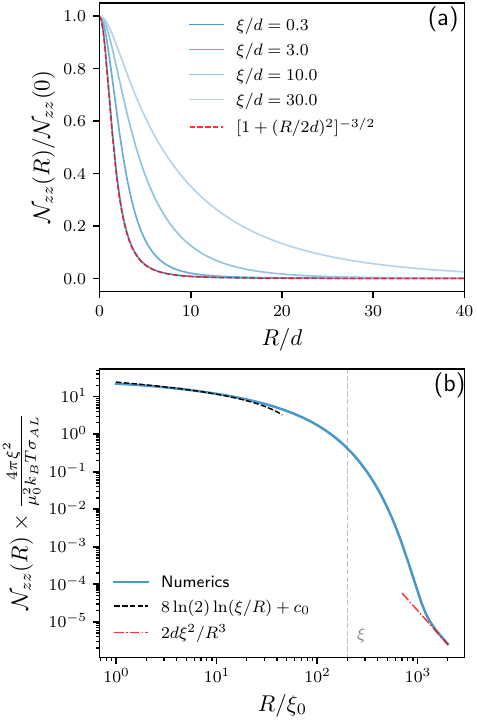} 
    \caption{(a) Normalized covariance $\mathcal N_{zz}(R)/\mathcal N_{zz}(0)$ at fixed distance $d$ for increasing $\xi/d$; the dashed curve is the exact local (metallic) form, Eq.~\eqref{eq:czzlocal}. The range of the covariance grows with the correlation length. (b) Deep in the fluctuation regime ($\xi = 200\,\xi_0$, $d = 0.25\,\xi_0$) the regimes of Eq.~\eqref{eq:czzregimes} are visible: logarithmic window (black dashed), and Ohmic tail $2d\xi^2/R^3$ (dash-dotted).}
    \label{fig:covariance}
\end{figure}

\textit{(ii) Fluctuation regime.} Inserting $\sigma'_T = \sigma^{\rm AL}F_T(q\xi/2)$ into Eq.~\eqref{eq:Nzz_Rzw}, the covariance profile develops additional structure once $\xi(T) \gg d$ (Fig.~\ref{fig:covariance}):
\begin{equation}
    \mathcal N^{\rm AL}_{zz}(R) \simeq \frac{\mu_0^2 k_BT\sigma^{\rm AL}}{4\pi} \times
    \begin{cases}
    \dfrac{8\ln 2}{\xi^2}\Big[\ln\dfrac{\xi}{R} + c_0\Big], & 2d \ll R \ll \xi, \\[1.5ex]
    \dfrac{2d}{R^3}, & R \gg \xi ,
    \end{cases}
    \label{eq:czzregimes}
\end{equation}
with $c_0$ an $\mathcal O(1)$ constant. In the window $2d \ll R \ll \xi(T)$ the covariance decays only logarithmically, with an amplitude that is independent of the reduced temperature (since $\sigma^{\rm AL}/\xi^2 = \gamma T e^2m/4\pi$). At $R \sim \xi(T)$ the profile is cut off and can be well-approximated by an expression involving modified Bessel functions (see Appendix~\ref{app:noise} for details). For the largest separations the profile always reverts to the universal Ohmic tail $\propto 2d\,\sigma^{\rm AL}/R^3$, generated by the linear-in-$q$ term of the evanescent kernel $e^{-2qd}$; its prefactor contains the full $1/\epsilon$ divergence of the paraconductivity. These regimes 
are verified numerically in Fig.~\ref{fig:covariance}. 

\section{Fluctuation diamagnetism}
\label{sec:diamagnetism}

The conductivity is the dissipative part of the fluctuation electromagnetic response; its reactive counterpart is the fluctuation diamagnetism \cite{schmid1969,prange1970,kurkijarvi1972,larkin2005}, known to be  large in the cuprates \cite{li2010}. It is natural to ask---particularly for an NV magnetometry experiment---whether fluctuating diamagnetic moments constitute an additional source of magnetic noise, and whether the diamagnetic response itself is detectable. To make more quantitative contact with experiment, in this section we also consider layered superconductors in the regime $\xi_c \ll a_c$, where $\xi_c$ is the coherence length in the perpendicular $c$-direction and $a_c$ is the $c$-axis lattice constant.  

The Gaussian fluctuation susceptibility of a 2D superconducting layer follows from the Landau-level spectrum of the pair fluctuations (Appendix~\ref{app:dia}); per layer,
\begin{equation}
    \chi_A(T) = -\frac{\pi}{3}\frac{k_BT\,\xi^2(T)}{\Phi_0^2} \;\propto\; -\frac1\epsilon,
    \label{eq:chiA}
\end{equation}
with $\Phi_0 = hc/2e_0$ the flux quantum. In a layered superconductor $1/\epsilon \to [\epsilon(\epsilon+r)]^{-1/2}$, with the Lawrence-Doniach anisotropy parameter $r = 4\xi^2_c(0)/a_c^2$.  
The fluctuation diamagnetism is thus exactly as singular as the paraconductivity---both are governed by $\xi^2(T)$---and, relative to its normal-state background (the Landau diamagnetism), it is parametrically much \textit{larger} than the conductivity correction relative to the Drude background \cite{larkin2005}. 
The nonlocal generalization $\chi(Q) = \chi_A\, G(Q\xi/2)$, with $G(0)=1$ and $G(x) \simeq 3\ln(2x)/x^2$ at $x\gg1$, was obtained in Ref. \cite{galaktionov1995}, whose thin-film kernel in fact contains our $F_T(\kappa,\varpi)$ as its dissipative part (Appendix~\ref{app:dia}).

\begin{figure*}[t!]
    \centering
    \includegraphics[width=\linewidth]{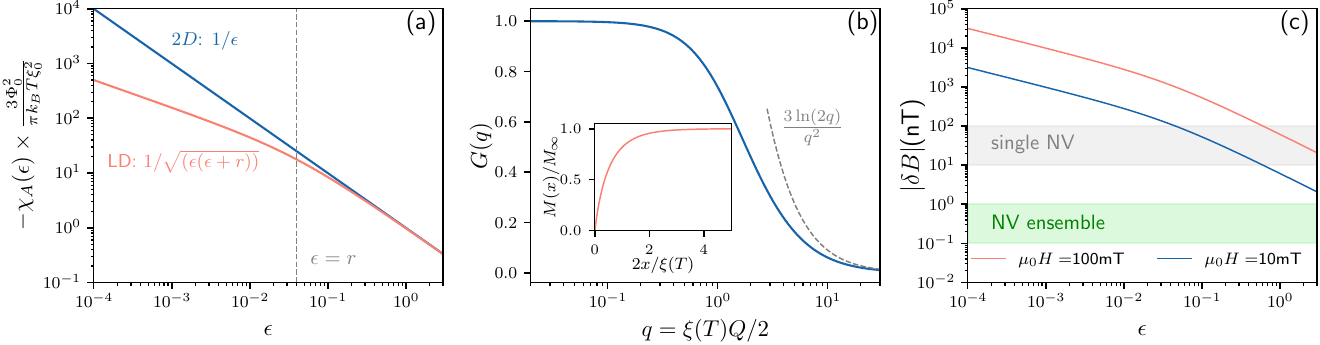}
    \caption{(a) Fluctuation susceptibility \eqref{eq:chiA}: 2D versus Lawrence-Doniach ($r = 0.04$). (b) Nonlocal scaling function $G(q)$ of $\chi(Q)$ \cite{galaktionov1995}, with the edge-healing profile $M(x)$ (inset). (c) Estimated edge stray field versus $\epsilon$ for the BSCCO flake geometry of Ref.~\cite{liu2025} at two applied fields, compared with single-NV and ensemble dc sensitivity bands.}
    \label{fig:diamagnetism}
\end{figure*}

\textit{(i) No new noise at zero field.---}A divergence-free sheet current and an out-of-plane magnetization density are the same degree of freedom: $\mathbf K = c\,\nabla\times(M_z\hat z)$ implies $K_T(\bfq) = icqM_z(\bfq)$, hence
\begin{equation}
    S_{K_TK_T}(\bfq,\omega) = c^2q^2\,S_{M_zM_z}(\bfq,\omega).
    \label{eq:nodouble}
\end{equation}
The equilibrium magnetic noise of the fluctuating diamagnetic loops is therefore \textit{identically} the transverse-current Johnson noise already encoded in $\sigma'_T(q,\omega)$: Eq.~\eqref{eq:Nzz_Rzw} contains all of it, and no double counting (or additional channel) arises at zero applied field. We have also checked that the noise generated at finite applied field by fluctuations of the local susceptibility ($\delta M = H\,\delta\chi$, sourced by $|\Psi|^2$ fluctuations) is negligible, smaller than the AL current noise by orders of magnitude at attainable fields.

\textit{(ii) Measurable response in an applied field.---}What is measurable is the diamagnetic response itself. In a perpendicular field $H$ the film acquires an areal magnetization $M_A = N_\ell\, \chi_A(T)\, H$ ($N_\ell$ = number of superconducting layers in the flake). A uniform infinite sheet produces no stray field, but edges, holes, and $T_c$ inhomogeneities do: near a straight edge the bound current $I = cM_A$ produces a stray field $|\delta B| \sim 2M_A/d$ at height $d$, with the edge profile healed over the length $\xi(T)/2$ \cite{galaktionov1995}. For the geometry of Ref.~\cite{liu2025} ($T_c = 90$~K, $\xi_0 \approx 2$~nm, flake thickness 200~nm so $N_\ell \approx 130$, $z = 50$~nm),
\begin{equation}
    |\delta B|_{\rm edge} \approx 
    \frac{6.3~\text{nT}}{\sqrt{\epsilon(\epsilon+r)}}\times\frac{\mu_0 H}{10~\text{mT}},
    \label{eq:dBedge}
\end{equation}
which reaches $\sim0.3~\mu$T at $\epsilon = 10^{-2}$ and remains above ensemble NV dc sensitivities (sub-nT) up to $\epsilon \sim 0.5$, i.e., tens of kelvin above $T_c$ (Fig.~\ref{fig:diamagnetism}). The measurement mode is the ESR line shift (dc magnetometry)---previously used for NV Meissner imaging \cite{bhattacharyya2024}---rather than relaxometry; field reversal $H\to-H$ flips the signal and provides clean lock-in background rejection; the response stays linear in $H$ up to $h \equiv H/H_{c2}(0) \sim \epsilon$. Beyond detection, spatial maps carry quantitative information: the edge-healing profile (Fig.~\ref{fig:diamagnetism}) measures $\xi(T)$ directly, and interior maps of $\chi_A(\epsilon(\mathbf r))$ can be used to image local-$T_c$ disorder.

\section{Nonequilibrium fluctuation noise}
\label{sec:noneq}

All results so far concern equilibrium noise, where the FDT ties the current fluctuations to the dissipative conductivity, Eq.~\eqref{current-fdt}. A dc electric field drives the fluctuation gas out of equilibrium. The nonlinear response was computed within TDGL by Dorsey \cite{dorsey1991} (following early work in Refs.~\cite{hurault1969,schmidtGor1968}) and beyond linear response the noise is no longer tied to the conductivity. Their difference is a direct, quantitative measure of the departure of the driven fluctuations from equilibrium. In this section we solve for the nonequilibrium current noise exactly---to all orders in the field and at all wave vectors---within Gaussian TDGL, and quantify the FDT violation. Details of the derivation are given in Appendix~\ref{app:noneq}.

\textit{(i) Exact solution---}In a uniform field ($\mathbf A = -c\mathbf Et$, $\mathbf E = E\hat x$) the linearized TDGL remains diagonal in the canonical momentum: each mode is an Ornstein-Uhlenbeck process with a ``sliding'' relaxation rate, $\gamma\partial_t\Psi_{\mathbf p} = -\varepsilon(\mathbf p + e\mathbf Et)\Psi_{\mathbf p} + \zeta_{\mathbf p}$. The two-time correlator is therefore exact [Eq.~\eqref{app:twotime}], with the driven mode occupation $N(\mathbf P) = \langle \Psi_{\mathbf p}\Psi^*_{\mathbf p} \rangle$ reproducing Eq.~(24) of Ref.~\cite{dorsey1991}. The mean current recovers Dorsey's nonlinear paraconductivity: in 2D, $\sigma(E) \equiv J/E = \sigma^{\rm AL}\Sigma_+(E/E_0)$ with $\Sigma_+(x) = \int_0^\infty \dd u\, e^{-u - x^2u^3}$ and the threshold field
\begin{equation}
    E_0 = \frac{\sqrt{12}\,\hbar}{2e\xi(T)\tau_{\rm GL}} \;\propto\; \epsilon^{3/2},
    \label{eq:E0}
\end{equation}
so that at criticality $\sigma(E) \propto E^{-2/3}$. We use the equivalent dimensionless field $f = 4\sqrt3\, E/E_0 = 2eE\xi\tau_{\rm GL}/\hbar$, the work done by the field across a coherence length during a GL lifetime.

Because the solution is Gaussian in $\mathbf P$ at fixed time arguments, Wick factorization reduces the symmetrized current noise at any wave vector to an explicit three-fold quadrature; see Appendix ~ \ref{app:noneq}, Eq.~\eqref{app:masternoneq}.

\begin{figure}[t!]
    \centering
    \includegraphics[width=\linewidth]{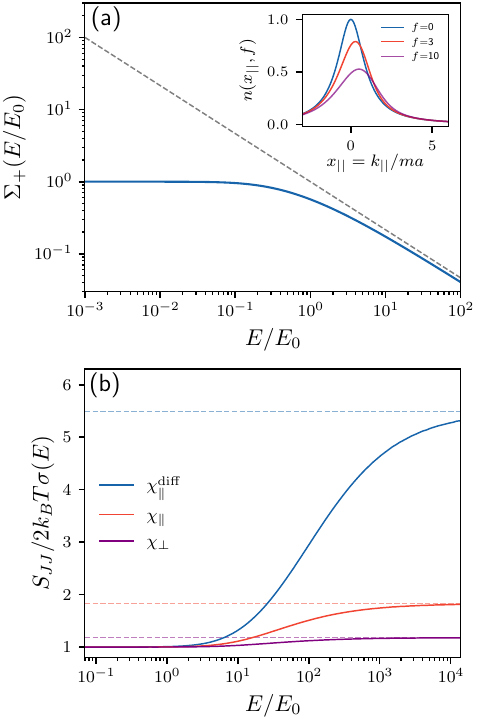}
    \caption{(a) Nonlinear AL conductivity $\Sigma_+(E/E_0)$ \cite{dorsey1991} with the critical $E^{-2/3}$ law; inset: the exact driven mode occupation $n(x_\parallel, f) = N(x_{\parallel}, f)/N(0, 0)$, whose drift skew sources the anisotropic noise. (b) FDT ratios $X = S_{JJ}(\omega\to0;E)/2k_BT\sigma(E)$ versus $E/E_0$; dotted lines mark the universal critical values, Eq.~\eqref{eq:universal}.}
    \label{fig:fdt}
\end{figure}

\textit{(ii) FDT breakdown at $k=0$---}Define the FDT ratios $X_\parallel = S_{xx}/2k_BT\sigma$, $X^{\rm diff}_\parallel = S_{xx}/2k_BT\sigma_{\rm diff}$ (with $\sigma_{\rm diff} = \dd J/\dd E$), and $X_\perp = S_{yy}/2k_BT\sigma$; for an isotropic film the transverse differential conductivity equals $\sigma_{\rm diff} = \sigma$ exactly, and all three ratios equal unity in equilibrium. We find (Fig.~\ref{fig:fdt}): at weak fields the FDT violation turns on quadratically,
\begin{equation}
    X_\parallel = 1 + c_\parallel\,(E/E_0)^2, \qquad X_\perp = 1 + c_\perp\,(E/E_0)^2,
\end{equation}
with constants $c_\parallel \approx 4.96$ and $c_\perp \approx 1.62$, 
while at strong fields---equivalently at criticality, where $\epsilon_{\rm eff}\sim(E/E_0)^{2/3}$---they saturate at \textit{universal numbers} of the 2D AL channel:
\begin{equation}
    X_\parallel \to X^{0}_\parallel \approx 1.83, \quad X^{\rm diff}_\parallel \to  
    3X^{0}_\parallel
    \quad X_\perp \to X_\perp^0 \approx 1.18,
    \label{eq:universal}
\end{equation}
the factor of 3 being exact since $J\propto E^{1/3}$ at $T_c$. The driven fluctuation gas is ``hotter'' than Johnson-Nyquist at the measured conductivity, with an effective noise temperature $T_{\rm eff} = X_\parallel^0 T$ (parallel) and $T_{\rm eff} = X_\perp^0 T$ (transverse). 
Physically, the field suppresses the response (a lifetime cutoff) more strongly than the noise, which weights the occupation squared; the noise therefore decays more slowly than $2k_BT\sigma(E)$.

\begin{figure}[t!]
    \centering
    \includegraphics[width=\linewidth]{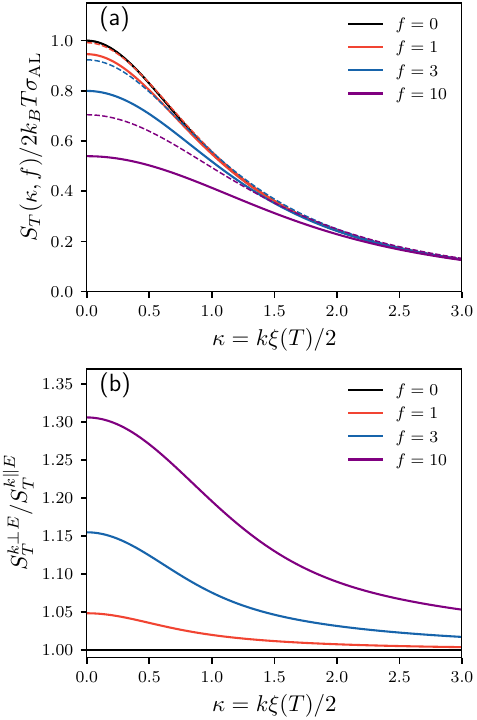}
    \caption{(a) Driven transverse noise $S_T(\kappa;E)$ for $\mathbf k\parallel\mathbf E$ (solid) and $\mathbf k\perp\mathbf E$ (dashed); dotted: equilibrium $F_T(\kappa)$. (b) The bias-induced anisotropy ratio.}
    \label{fig:noneqk}
\end{figure}

\textit{(iii) Bias-induced noise anisotropy at finite $k$---}At finite wave vector the driven transverse noise splits by orientation (Fig.~\ref{fig:noneqk}): fluctuations with $\mathbf k \parallel \mathbf E$ are suppressed more strongly than those with $\mathbf k \perp \mathbf E$, with the anisotropy ratio reaching $\sim1.05$, $1.15$, and $1.30$ at $f = 1$, $3$, and $10$ for $\kappa \lesssim 0.5$, and closing at $\kappa \gtrsim 2$, where short-wavelength modes are stiffer than the drift scale. This is precisely the geometry that covariance magnetometry resolves: two NV sensors separated parallel versus perpendicular to an applied bias current measure different covariances, with a contrast of tens of percent at $f \sim 1$---a clear signature of nonequilibrium superconducting fluctuations. Combined with a conventional measurement of $\sigma(E)$ on the same device, the NV noise provides a direct experimental test of the FDT violation, Eq.~\eqref{eq:universal}. We note that a complementary covariance signature---anisotropic noise from vortex drift below $T_c$---was proposed recently in Ref.~\cite{zhang2026}; the present effect lives above $T_c$, where the theory is fully controlled.

\textit{(iv) Why heating does not spoil the measurement---}One might worry that applying a strong electric field simply leads to Joule heating, making the problem ill-defined. The key observation, however, is that the threshold field for the nonlinear response of the fluctuation Cooper pairs, Eq.~\eqref{eq:E0}, collapses rapidly as $T\to T_c$, $E_0 \propto (T-T_c)^{3/2}$, whereas the nonlinearity (and heating) scales of the normal quasiparticle fluid are set by microscopic energies and are temperature independent in this window. Consequently, there is a parametrically broad regime in which the fluctuation contribution is driven deep into the nonlinear, non-equilibrium regime while the normal-state response remains strictly linear and the quasiparticle bath remains at the lattice temperature: the assumption of a fixed bath temperature $T$ underlying the TDGL Langevin dynamics is then self-consistent, and the fluctuation correction is cleanly separable, as it is precisely the part of the signal with the anomalous field, temperature, and wave-vector dependence derived above. Quantitatively, $f = 1$ corresponds to $E \approx 25$~V/cm for BSCCO at $\epsilon = 10^{-2}$ (dropping as $\epsilon^{3/2}$ closer to $T_c$), i.e., sheet current densities of a few A/m per layer---accessible with standard pulsed-bias techniques on narrow bridges, with duty cycling suppressing the average dissipated power. This opens the realistic possibility of experimentally probing the breakdown of the fluctuation-dissipation theorem through measurements of nonequilibrium current noise in the Cooper-pair fluctuation channel.

\section{Summary of main results}
\label{sec:summary}

For convenience, we collect the main results of this work.

\textit{(i) Nonlocal paraconductivity.} Within TDGL, the nonlocal AL paraconductivity has the closed form $\sigma_{ij} = \sigma^{\rm AL}[(\delta_{ij}-\hat\kappa_i\hat\kappa_j)F_T + \hat\kappa_i\hat\kappa_j F_L]$, with the $d=2$ scaling functions of Eq.~\eqref{eq:ft} and their finite-frequency generalization [Appendix~\ref{app:AL}, Eq.~\eqref{app:eq_FTkw}]; the kernel agrees with (and extends the observables associated to) the thin-film fluctuation electrodynamics of Refs.~\cite{barash1993,galaktionov1995}.

\textit{(ii) Single-NV relaxometry.} The critical enhancement $1/T_1 \propto 1/\epsilon$ observed in Ref.~\cite{liu2025} follows from the 2D AL conductivity with mean-field exponent unity; it is rounded off below \textit{two} distinct scales---the distance scale $\epsilon_d \sim (\xi_0/d)^2$ (Eq.~\eqref{eq:T1limits}) and the frequency scale $\epsilon_\omega \sim \hbar\omega/k_BT_c$
---so that frequency-resolved relaxometry measures the critical slowing down $\tau_{\rm GL}(\epsilon)$.

\textit{(iii) Covariance magnetometry.} The two-point field correlator, Eq.~\eqref{eq:Nzz_Rzw}, is a wave-vector-resolved probe: geometric and temperature independent in the metal (Eq.~\eqref{eq:czzlocal}), it develops the structure of Eq.~\eqref{eq:czzregimes} in the fluctuation regime, with a range that measures $\xi(T)$---potentially enabling, e.g., a model-independent discrimination between Gaussian ($\xi\propto\epsilon^{-1/2}$) and BKT (exponential $\xi(T)$) fluctuation regimes.

\textit{(iv) Spin channel.} The anomalous MT correction to the spin susceptibility, Eq.~\eqref{eq:MTchi}, produces a relaxation enhancement $\propto \ln(\epsilon/\gamma_\phi)/(\epsilon-\gamma_\phi)$ cut off by pair breaking, and a covariance of range $L_\phi$ (Eq.~\eqref{eq:Cspin}), separable from the orbital channel by its temperature dependence; for $d$-wave BSCCO this channel is suppressed, identifying the observed noise peak as orbital.

\textit{(v) Fluctuation diamagnetism.} At zero field, magnetization noise is not an additional channel (Eq.~\eqref{eq:nodouble}); in an applied field, the diamagnetic response produces edge stray fields of order $0.01$--$1~\mu$T (Eq.~\eqref{eq:dBedge}), measurable by NV dc magnetometry tens of kelvin above $T_c$ and yielding local maps of $\xi(T)$ and $T_c$ disorder.

\textit{(vi) Nonequilibrium noise.} The current noise of the driven fluctuation gas is obtained exactly to all orders in the bias and at all wave vectors (Eq.~\eqref{app:masternoneq}); it violates the FDT by the universal critical ratios of Eq.~\eqref{eq:universal} and acquires a bias-induced spatial anisotropy of tens of percent---both measurable with NV sensors, with the threshold field $E_0\propto\epsilon^{3/2}$ making the nonlinear fluctuation regime accessible at modest current densities while the normal fluid remains Ohmic.

The exact result obeys the finite-temperature critical scaling form
$S_{\mu\nu}=2k_BT\sigma^{\rm AL}(\epsilon)\,\mathcal S_{\mu\nu}(k\xi,\omega\xi^z,E\xi^{1+z})$ with the
mean-field exponents $\nu=1/2$ and $z=2$ of relaxational dynamics, the electric field entering only
through the combination $E\xi^{1+z}$. In the critical limit the noise reduces to
$S=X^*_\parallel\,2k_BT\,\sigma(E)$ with $\sigma(E)\propto E^{-2/3}$ and spectral weight collapsing on the scale $\tau_E^{-1}\propto E^{z/(1+z)}$---the thermal, $z=2$ counterpart of the universal
nonequilibrium noise scaling derived for the $z=1$ superconductor--insulator quantum critical point \cite{Green2006}, whose form $S_j=T\Phi[T_{\rm eff}(E)/T]$,
$T_{\rm eff}\propto E^{z/(1+z)}$, is recovered here with the classical identification
$T_{\rm eff}=X^*T$. Logarithmic factors in our expressions are confined to the asymptotics of the
scaling functions and do not modify the exponents.

\section{Discussion}
\label{sec:discussion}

We have analyzed the contribution of SC fluctuations to the relaxation rate of single NV centers and covariance measurements of spatially separated NVs. In the case of an individual NV sensor, we have extended the theory presented in Ref.~\cite{liu2025} to account for the nonlocal, frequency-dependent paraconductivity, which determines the relaxation rate as a function of tip-to-sample distance $d$ and NV frequency $\omega$. The wave vector dependence of the paraconductivity is especially important in the regime $d < \xi(T)$. In the case of covariance measurements, we have shown how the nonlocal paraconductivity determines the scaling of the covariance with distance $R$ between NVs; in the interesting regime $d\ll R \ll \xi(T)$, we find a slow logarithmic decay with temperature-independent amplitude, terminated by an exponential shoulder at $R\sim\xi(T)$ whose decay length is $\xi(T)/2$, and ultimately by the universal Ohmic tail $\propto 2d\,\sigma^{\rm AL}/R^3$.

What new information about the superconducting state do these measurements provide? At the Gaussian level the AL noise is blind to the pairing symmetry---it measures lengths and times: $\xi(T)$ and its exponent $\nu$ (through the covariance range), $\tau_{\rm GL}(\epsilon)$ and the dynamical exponent $z_{\rm dyn}$ (through the frequency cutoff $\epsilon_\omega$), and the dephasing time $\tau_\phi$ (through the saturation of the MT spin channel). These are precisely the quantities that distinguish competing scenarios for two-dimensional superconductivity: a Gaussian correlation length $\xi\propto\epsilon^{-1/2}$ versus the exponential BKT divergence versus 3D-XY criticality; relaxational versus propagating order-parameter dynamics; and intrinsic critical correlations versus static inhomogeneity, which produce covariance ranges with sharply different temperature dependence. Pairing symmetry enters indirectly but usefully: strong pair breaking in a $d$-wave superconductor quenches the anomalous MT channel, and the diamagnetic response of Sec.~\ref{sec:diamagnetism} adds an independent, response-based observable with its own $\xi(T)$ content. Out of equilibrium, the universal FDT-violation ratios, Eq.~\eqref{eq:universal}, characterize the driven pair-fluctuation gas itself and have no analogue in transport.

It is important to note that the Gaussian AL theory is controlled only when the fluctuation corrections are small compared to the normal state conductivity. In the experiments \cite{liu2025}, the NV relaxation rate was found to increase by roughly an order of magnitude near $T_c$. Per CuO$_2$ layer, the Gaussian expectation is $(1/T_1)_{\rm fluct}/(1/T_1)_0 = R_\square e_0^2/16\hbar\epsilon \approx 0.015/\epsilon$ for a normal-state sheet resistance $R_\square\approx1$~k$\Omega$, reaching a tenfold enhancement only at $\epsilon\sim10^{-3}$---far narrower than the observed few-kelvin peak. A more sophisticated theoretical treatment is thus required to explain the magnitude of this effect, taking into account the critical (BKT) nature of the 2D SC transition \cite{curtis2024} and, within the Ginzburg window, the self-consistent (Hartree) renormalization of the pair propagator \cite{ullah1991}. We also note that the present framework can be straightforwardly extended to include the effects of an out-of-plane, orbital magnetic field by working in the Landau-level basis---a direction of immediate experimental relevance, given the field-dependent data already reported in Ref.~\cite{liu2025}, and one that connects naturally to the diamagnetic response of Sec.~\ref{sec:diamagnetism}.

In this paper, we have only considered SC fluctuation effects above $T_c$. Below $T_c$, the situation is more complex, as the noise is expected to receive contributions from both the amplitude and phase fluctuations of the order parameter, as well as collective modes \cite{carlson1975,schmidschon1975}. It will be interesting to understand whether nonlocal NV covariance magnetometry can shed light on the problem of collective modes in SCs.

\textit{Note added.---}During the final stages of this work we became aware of a recent preprint by Orgad \cite{orgad2026}, which addresses a closely related problem. Where the two works overlap---the equilibrium two-point (covariance) noise spectra generated by Gaussian superconducting fluctuations, the current noise to lowest order in an applied bias (linear order in the Ref.~\cite{orgad2026} due to broken particle-hole symmetry), and bias-induced covariance anisotropy---we find complete agreement. The present work extends the analysis in several additional directions, including the Maki-Thompson spin channel and its covariance (Sec.~\ref{sec:spin}), the fluctuation-diamagnetism response channel (Sec.~\ref{sec:diamagnetism}), the exact all-order nonequilibrium current noise with its universal fluctuation-dissipation-violation ratios (Sec.~\ref{sec:noneq}), and the quantitative connection to the experiments of Ref.~\cite{liu2025}.

\begin{acknowledgments}

We thank Shubhayu Chatterjee for a discussion of the data in Ref.~\cite{liu2025} and Andrey Varlamov for pointing our attention to Ref. \cite{barash1993}. We are particularly grateful to Dror Orgad for valuable comments and for communicating with us regarding Ref. \cite{orgad2026}. This work was supported by the U.S. Department of Energy (DOE), Office of Science, Basic Energy Sciences (BES) under Award No. DE-SC0020313.
A. L. acknowledges H. I. Romnes Faculty Fellowship provided by the University of Wisconsin-Madison Office of the Vice Chancellor for Research and Graduate Education with funding from the Wisconsin Alumni Research Foundation.
G.R. acknowledges financial support from the Sweden-America Foundation through the Ingegerd \& Viking Olov Bj\"orks stipendiefond.
The authors acknowledges the use of Claude (Anthropic) \cite{Claude2026} with manuscript preparation. 

\end{acknowledgments}

\bibliography{noise_sc_fluct}
\begin{widetext}
\appendix

\section{Magnetic noise tensor}
\label{app:noise}
The magnetic noise tensor $\mathcal{N}_{ab} = \langle B_a(\mathbf r_1, z\,;t_1)B_b(\mathbf r_2, z\,;t_2) \rangle$ can be calculated directly using the Biot-Savart (BS) kernel,
\begin{equation}
    B_a(\mathbf r,t) = \int g_{ab}(\mathbf r-\mathbf r')J_b(\mathbf r', t)\dd ^3\mathbf r'.
\end{equation}
Using the fluctuation dissipation theorem we relate the noise to the response and using mode expansions we find,
\begin{equation}
    \mathcal{N}_{ab}(\mathbf R, z, \omega) = \coth\big(\frac{\omega\beta}{2}\big)\int e^{i\mathbf R\cdot \mathbf q}g_{aa'}(\mathbf q)g_{bb'}(-\mathbf q)\Im\chi_{a'b'}^J(\mathbf q, \omega).
\end{equation}
The BS kernel can be found directly from Maxwell's equations in the non-relativistic limit using the London gauge $\nabla\cdot \mathbf A =0$,
\begin{equation}
    (q^2 - \partial^2_z)\mathbf A(\mathbf q, z) = \mu_0\mathbf J(\mathbf q)\delta(z)
\end{equation}
Solving this inhomogeneous DE in $z$ and taking the curl we find,
\begin{equation}
    B_a(\mathbf q, z) = g_{ab}(\mathbf q,z)J_b(\mathbf q), \quad g_{ab}(\mathbf q, z) = -\frac{\mu_0}{2}e^{-qz}[(i\hat{\mathbf z} + \hat{\mathbf q})\times \hat e_a]\cdot\hat e_{b}
\end{equation}
The current response is related to conductivity by $\chi^J = i\omega \sigma$ and decomposing the conductivity tensor into the transverse and longitudinal parts we find, in the classical limit $\coth(\omega\beta/2)\to 2k_BT/\hbar\omega$ [the two-sided symmetrized convention, $S_{K_TK_T} = 2k_BT\sigma'_T$, consistent with Eq.~\eqref{current-fdt}],
\begin{align}
    \mathcal{N}_{zz}(\mathbf R, z,
    \omega) &= \frac{\mu_0^2k_BT}{4\pi}\int_0^\infty\dd q\, qJ_0(Rq)e^{-2qz}\sigma'_T(q,\omega), \label{app:Nzz_Rzw}\\
    \mathcal{N}_{xx}(\mathbf R, z, \omega) &= \frac{\mu_0^2k_BT}{4\pi R}\int_0^\infty \dd q\,  J_1(Rq)e^{-2qz}\sigma'_T(q, \omega) .
\end{align}
The overall normalization can be checked against the thin-film limit of the reflection-coefficient formulation of Ref.~\cite{dolgirev2022}, with which it agrees. The $xx$ component depends on the longitudinal part of the conductivity only through relativistic corrections and is dominated by $\sigma_T$ in the non-relativistic limit. The limit $R\to0$ recovers the single NV equations used in the main text,
\begin{align}
    \mathcal{N}_{zz}(
    \omega) &= \frac{\mu_0^2k_BT}{4\pi}\int_0^\infty\dd q\, qe^{-2qz}\sigma'_T(q,\omega), \label{app:Nzz_w}\\
    \mathcal{N}_{xx}(\omega) &= \frac{\mu_0^2k_BT}{8\pi}\int_0^\infty \dd q\, q e^{-2qz}\sigma'_T(q, \omega),
\end{align}
so that $\mathcal N_{xx} = \mathcal N_{yy} = \tfrac12 \mathcal N_{zz}$. The NV relaxation rate is set by the field noise transverse to the NV axis $\hat n$, $1/T_1 = (g\mu_B/\hbar)^2\, [\,\mathrm{Tr}\,\mathcal N - \hat n\cdot\mathcal N\cdot\hat n\,]$, which for the axis at angle $\theta$ from the film normal gives $1/T_1 = (g\mu_B/\hbar)^2\, s_{\hat n}\,\mathcal N_{zz}$ with $s_{\hat n} = (3-\cos^2\theta)/2$ \cite{casola2018probing}; additional details regarding NV relaxation rates can be found in, e.g., \cite{cambria2021,cambria2023}.

As discussed in Appendix~\ref{app:AL} the conductivity is conveniently written $\sigma_T(\kappa, \varpi) = \sigma_{AL}F_T(\kappa, \varpi)$ and we introduce the notation,
\begin{equation}
    \mathcal{N}_{zz}(\mathbf R, z,
    \omega=0) = \frac{\mu_0^2k_BT\sigma_{AL}}{4\pi}\int_0^\infty\dd q ~ q e^{-2dq}J_0(Rq)F_T\left(\frac{\xi q}{2}\right).
\end{equation}

\subsection{Limiting behaviors of the noise}
Focusing on the static case $\omega = 0$, we can derive the behavior of $\mathcal N_{zz}$ in the different regimes $d\gg \xi$ and $d \ll \xi$. The $d \gg \xi$ is elementary and done in the main text, so let us focus on the more interesting regime $d \ll \xi$. We focus on the integral
\begin{equation}
    \tilde{\mathcal N}_{zz} = \int_0^\infty \dd q ~ q e^{-2dq}J_0(Rq)F_T(\xi q/2) = \frac{4}{\xi^2}\int_0^\infty \dd x ~ xe^{-\frac{4d}{\xi}}J_0(2Rx/\xi)F_T(x).
\end{equation}
Since we are interested in the regime $\xi \gg d$, we can neglect the exponential factor and investigate
\begin{equation}
    I(\alpha) = \int_0^\infty \dd x ~ x J_0(\alpha x) F_T(x)
\end{equation}
with $\alpha =2R/\xi$. Due to the oscillatory nature of $J_0(\alpha x)$ we will analytically continue the argument into the complex plane where the oscillations turn into damping. We employ $J_0(x) = [H_{0}^{(1)}(x) + H_0^{(2)}(x)]/2$ to split the integral $I$ into two parts $I=I_++I_-$,  where $I_+$ contains $H_0^{(1)}$ and $I_-$ contains $H_0^{(2)}$. $I_+$ can be continued into the upper half-plane via an infinite quasi-circle covering the 1st quadrant and $I_-$ can be continued into the lower half-plane the same way but for the 4th quadrant. It is easy to see that
\begin{align}
    I_+(\alpha) &= -\frac12\int_0^\infty \dd x\, xH_0^{(1)}(iRx)F_T(ix + 0^+),\\
    I_-(\alpha) &= -\frac12\int_0^\infty \dd x\, x H_0^{(2)}(-iRx)F_T(-ix + 0^+).
\end{align}
The infinitesimal real part in the argument of $F_T$ is included to avoid the poles on the imaginary axis. Using $H_0^{(1)}(ix) = -H_0^{(2)}(-ix) = 2K_0(x)/(i\pi)$ we find $I_- = I_+^\ast$ and
\begin{equation}
    I = I_+ + I_+^\ast = 2\text{Re} I_+ = -\frac2\pi\int_0^\infty \dd x K_0(\alpha x)\text{Im} F_T(ix + 0^+).
\end{equation}
The finite $d$ case can be restored here by $F_T \to \exp(-i4dx/\xi)F_T$ but the expressions become too cumbersome to write here. Looking at the analytic continuation 
\begin{equation}
    F_T(ix+0^+) = \begin{cases}
        \dfrac{2\arcsin(x)}{x\sqrt{1-x^2}} + \dfrac{\ln(1-x^2)}{x^2},\quad  &x<1,\\
        -2\dfrac{\text{arccosh}(x)}{x\sqrt{x^2-1}} + \dfrac{\ln(1-x^2)}{x^2} + i\pi \left(\dfrac{1}{x^2} - \dfrac{1}{x\sqrt{x^2-1}}\right), \quad &x>1,
    \end{cases}
\end{equation}
we see that the imaginary part of $F_T$ is zero below $x=1$ and hence
\begin{equation}\label{eq:analytic_expression_fluctuation_limit}
    I(\alpha) = 2\int_1^\infty \dd x\, K_0(\alpha x) \left(\frac{1}{\sqrt{x^2-1}}-\frac1x\right) = [K_0(\alpha/2)]^2 - 2\int_1^\infty\dd x\,\frac{K_0(\alpha x)}x.
\end{equation}
This form is exact for $d=0$ and a good approximation for $d \ll R \lesssim\xi$. The true $R\gg \xi\gg d$ limit cannot be restored with Eq.~\ref{eq:analytic_expression_fluctuation_limit} since the finite $d$ eventually restores the Ohmic tail shown in Fig.~\ref{fig:covariance}. Let us instead look at the limit $\alpha \ll 1$ corresponding to $\xi\gg R\gg d$. Since $K_0(x\ll 1)\approx \ln(2/x) - \gamma$, 
\begin{equation}
    \xi^2\tilde{\mathcal{N}}_{zz}(\xi\gg R\gg d) = 8\ln(2)\left[\ln\left(\frac{\xi}{R}\right) - c_0\right],
\end{equation}
matching the form in Eq.~\ref{eq:czzregimes} and Fig.~\ref{fig:covariance}. The constant $c_0$ is,
\begin{equation}
    c_0 = \gamma + \frac{\pi^2}{48\ln2} - \frac{\ln2}{4}. 
\end{equation}
For small, nonzero $d$, $c_0$ can be treated as a fitting parameter. 

\section{Derivation of AL contribution to conductivity}
\label{app:AL}
Starting from the expression of the Eq.~\eqref{ConductivityTensor} we perform the integral over frequency by methods of residues and find,
\begin{equation}
\sigma'_{\mu\nu}(\mathbf{k},\omega)=\frac{\gamma
Te^{2}}{m^{2}}\int p_{\mu}p_{\nu}
\frac{\varepsilon_{\mathbf{p}+\mathbf{k}/2}+
\varepsilon_{\mathbf{p}-\mathbf{k}/2}}{\varepsilon_{\mathbf{p}+\mathbf{k}/2}
\varepsilon_{\mathbf{p}-\mathbf{k}/2}[(\varepsilon_{\mathbf{p}+\mathbf{k}/2}
+\varepsilon_{\mathbf{p}-\mathbf{k}/2})^{2}+\gamma^{2}\omega^{2}]}\frac{\mathrm{d}^{d}p}{(2\pi)^{d}}
\end{equation}
The imaginary part can be restored using Kramers-Kronig relation,
\begin{equation}
\sigma''_{\mu\nu}(\mathbf{k},\omega)=\frac{\gamma
Te^{2}}{m^{2}}\int p_{\mu}p_{\nu}
\frac{\gamma\omega}{\varepsilon_{\mathbf{p}+\mathbf{k}/2}
\varepsilon_{\mathbf{p}-\mathbf{k}/2}[(\varepsilon_{\mathbf{p}+\mathbf{k}/2}
+\varepsilon_{\mathbf{p}-\mathbf{k}/2})^{2}+\gamma^{2}\omega^{2}]}\frac{\mathrm{d}^{d}p}{(2\pi)^{d}}.
\end{equation}
Combining real and imaginary parts together, introducing
dimensionless momenta ${\vec \rho}=\mathbf{k}\xi$, ${\vec
\kappa}=\mathbf{k}\xi/2$ and dimensionless frequency
$\varpi=\omega/\Omega$, where $\Omega=1/\gamma m\xi^{2}$ we find
conductivity tensor
$\sigma_{\mu\nu}(\mathbf{k},\omega)=\sigma_{AL}F_{\mu\nu}(\mathbf{k},\omega)$ with,
\begin{equation}
F_{\mu\nu}(\kappa,\varpi)=\frac{4(4\pi)^{d/2}}{\Gamma(2-d/2)}\int\frac{\rho_{\mu}\rho_{\nu}}
{[1+\rho^{2}+\kappa^{2}+2{\vec\kappa}{\vec\rho}]
[1+\rho^{2}+\kappa^{2}-2{\vec\kappa}{\vec\rho}]
[1+\rho^{2}+\kappa^{2}-i\varpi]}
\frac{\mathrm{d}^{d}\rho}{(2\pi)^{d}}.
\end{equation}
The momentum integral can be performed by Feynman parametrization,
\begin{equation}
    \frac{1}{ab} = \int_0^1\frac{\dd x}{(ax+(1-x)b)^2}, \quad \frac{1}{a^2b} = -\frac{\partial}{\partial a} \frac{1}{ab} = \int_0^1\frac{2x\dd x}{(ax+(1-x)b)^3}
\end{equation}
Inserting these identities and completing the square in the denominator we find,
\begin{equation}\label{ScalingTensorIntegral-2}
F_{\mu\nu}(\kappa,\varpi)=\frac{8(4\pi)^{d/2}}{\Gamma(2-d/2)}
\int\limits_{0}^{1}\mathrm{d}z\int\limits_{0}^{1}\mathrm{d}y\
y\int\frac{\mathrm{d}^{d}q}{(2\pi)^{d}}\frac{q_{\mu}q_{\nu}+z^{2}y^{2}\kappa_{\mu}\kappa_{\nu}}
{[1+q^{2}+\kappa^{2}(1-z^{2}y^{2})-(1-y)i\varpi]^{3}}
\end{equation}
After integration over the momentum $q$ we find following formulas
for the transverse and longitudinal scaling functions
\begin{equation}
F_{T}(\kappa,\varpi)=\int\limits_{0}^{1}\int\limits_{0}^{1}
\frac{2y\mathrm{d}z\mathrm{d}y}{[1+\kappa^{2}(1-z^{2}y^{2})-(1-y)i\varpi]^{2-d/2}}
\end{equation}
\begin{equation}
F_{L}(\kappa,\varpi)=F_{T}(\kappa,\varpi)+(2-d/2)\kappa^{2}
\int\limits_{0}^{1}\int\limits_{0}^{1}
\frac{z^{2}y^{3}\mathrm{d}z\mathrm{d}y}{[1+\kappa^{2}(1-z^{2}y^{2})-(1-y)i\varpi]^{3-d/2}}
\end{equation}
For $d=2$ the transverse integral can be done exactly. Making the change in variables $x=zy$ the domain is $0\le x\le y$ and the integral becomes,
\begin{equation}
    F_T(\kappa, \varpi) = \int_0^1\dd y\int_0^y \dd x \frac{2}{1+\kappa^2(1-x^2) -(1-y)i\varpi}.
\end{equation}
Now changing order of integration ($x\le y\le 1$) and integrating w.r.t. $y$ we find,
\begin{equation}
    F_T(\kappa, \varpi) = \frac{2}{i\varpi}\int_0^1\dd x \ln\left(\frac{1+\kappa^2(1-x^2)}{1+\kappa^2(1-x^2)-(1-x)i\varpi}\right)
\end{equation}
This is now just a sum of two integrals of the form,
\begin{equation}
    I= \int_0^1\dd x \ln(ax^2+bx+c),
\end{equation}
Performing these integrals and simplifying we find,
\begin{equation}
    F_T(\kappa, \varpi) = \frac{2}{i\varpi}\left[\frac{2\sqrt{1+\kappa^2}}{\kappa}\mathrm{arcsinh}(\kappa) - \ln(-\kappa^2) -\sum_{\sigma=\pm}\big((1-r_\sigma)\ln(1-r_\sigma)+r_\sigma\ln(-r_\sigma)\big)\right]
\end{equation}
with $\ln$ being the principal value logarithm $\ln(-1) \equiv i\pi$ and,
\begin{equation}
    r_\pm = \frac{i\varpi}{2\kappa^2}\pm \sqrt{1+\frac{1}{\kappa^2}-\frac{i\varpi}{\kappa^2} +\frac{(i\varpi)^2}{4\kappa^4}}.
\end{equation}
This can be simplified to the form,
\begin{align}\label{app:eq_FTkw}
    F_T(\kappa, \varpi) &= \frac{2}{i\varpi}\left[\frac{2\sqrt{1+\kappa^2}}{\kappa}\mathrm{arcsinh}(\kappa)-\frac2\kappa g(\kappa, \varpi)\mathrm{arctanh}\left(\frac{2\kappa g(\kappa,\varpi)}{2+2\kappa^2-i\varpi}\right)\right] - \frac{\ln(1+\kappa^2-i\varpi)}{\kappa^2},\\
    g(\kappa, \varpi) &= \sqrt{1+\kappa^2-i\varpi-\varpi^2/4\kappa^2}
\end{align}
Expanding this expression for $\varpi\to0$ one easily finds,
\begin{equation}
    F_T(\kappa, \varpi\to0) = \frac{2\mathrm{arcsinh}(\kappa)}{\kappa\sqrt{1+\kappa^2}} -\frac{\ln(1+\kappa^2)}{\kappa^2}
\end{equation}
As a sanity check, Eq.~\eqref{app:eq_FTkw} has been numerically verified to satisfy Kramers-Kronig relations, to agree with direct numerical evaluation of the Feynman-parametric representation, and to coincide with the thin-film response kernel of Ref.~\cite{galaktionov1995} through the identity $B_G(\kappa,\varpi) = B_G(\kappa,0) + \tfrac{i\varpi}{4}F_T(\kappa,\varpi)$, where $B_G$ denotes the bracket of Eq.~(6) of that reference and $B_G(\kappa,0) = 1 - \sqrt{1+\kappa^2}\,\mathrm{arcsinh}(\kappa)/\kappa$ is its static (diamagnetic) part; see Appendix~\ref{app:dia}.

\begin{figure}[b]
    \centering
    \includegraphics[width=\linewidth]{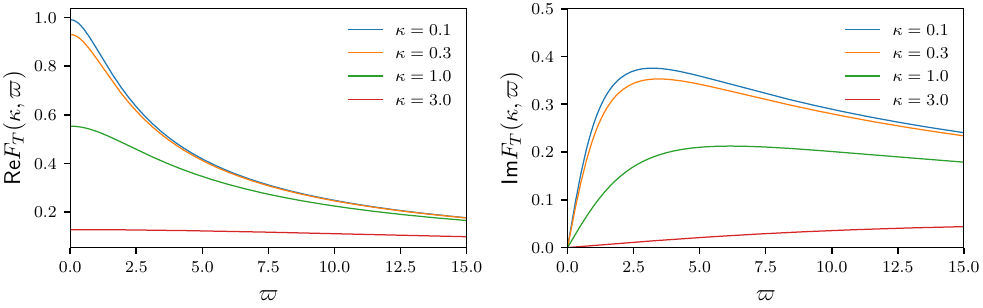}
    \caption{Real and imaginary parts of the dynamical scaling function $F_T(\kappa,\varpi)$, Eq.~\eqref{app:eq_FTkw}, versus the dimensionless frequency $\varpi = \omega\gamma m\xi^2$ for several values of $\kappa = k\xi/2$. The dissipative part is flat for $\varpi \lesssim 1$ (white-noise regime relevant to NV relaxometry at $\epsilon \gg \epsilon_\Omega$) and falls off at $\varpi \gtrsim 1$, where the probe frequency exceeds the fluctuation relaxation rate; the reactive part peaks at the crossover. }
    \label{fig:conddyn}
\end{figure}

\section{Nonequilibrium current noise: exact solution}
\label{app:noneq}

In a uniform electric field, $\mathbf A = -c\mathbf E t$ with $\mathbf E = E\hat x$, the linearized TDGL equation remains diagonal in the canonical momentum $\mathbf p$, with the kinetic momentum sliding as $\mathbf P(t) = \mathbf p + e\mathbf Et$. Each mode is an Ornstein-Uhlenbeck process with time-dependent rate, and the two-time correlator is exact: for $\tau = t-t' \ge 0$, with $\mathbf P$ the kinetic momentum at the earlier time,
\begin{equation}
    \langle \Psi(t)\Psi^*(t')\rangle_{\mathbf p} = \exp\left[-\frac1\gamma\int_0^\tau \varepsilon(\mathbf P + e\mathbf E u)\,\dd u\right] N(\mathbf P), \qquad
    N(\mathbf P) = \frac{2T}{\gamma}\int_0^\infty \dd s\; e^{-\frac2\gamma \int_0^s \varepsilon(\mathbf P - e\mathbf E v)\,\dd v},
    \label{app:twotime}
\end{equation}
where $N(\mathbf P)$ is the driven occupation [equal to $T/\varepsilon_{\mathbf P}$ at $E=0$, and reproducing Eq.~(24) of Ref.~\cite{dorsey1991}]. The mean current $\mathbf J = (e/m)\int_{\mathbf P}\mathbf P N(\mathbf P)$ recovers the nonlinear AL conductivity, $\sigma(E) = \sigma^{\rm AL}\Sigma_+(E/E_0)$ \cite{dorsey1991}.

Since the theory is Gaussian, Wick factorization gives the symmetrized zero-frequency current noise at any wave vector as
\begin{equation}
    S_{\mu\nu}(\mathbf k, \omega\to0; E) = 2\left(\frac em\right)^2 \int \frac{\dd^2P}{(2\pi)^2}\int_0^\infty \dd\tau\; (\mathbf P + e\mathbf E\tau)_\mu P_\nu\, N\big(\mathbf P + \tfrac{\mathbf k}2\big) N\big(\mathbf P - \tfrac{\mathbf k}2\big)\, e^{-\frac1\gamma\int_0^\tau\left[2\varepsilon(\mathbf P + e\mathbf Eu) + \frac{k^2}{4m}\right]\dd u},
    \label{app:Sphys}
\end{equation}
where the identity $\varepsilon(\mathbf P + \tfrac{\mathbf k}2) + \varepsilon(\mathbf P - \tfrac{\mathbf k}2) = 2\varepsilon(\mathbf P) + k^2/4m$ has been used. All time integrals carry cubic-in-time damping from the field [$\int_0^s(\ldots + e\mathbf Ev)^2\dd v \supset e^2E^2s^3/3$], which regularizes the $\epsilon\to0$ limit and generates the threshold field Eq.~\eqref{eq:E0}. Inserting the integral representation of $N$ and performing the exact Gaussian $\dd^2P$ integral reduces Eq.~\eqref{app:Sphys} to a three-fold quadrature with explicit normalization. With $\kappa = k\xi/2$, $\theta$ the angle between $\mathbf k$ and $\mathbf E$, $f = 4\sqrt3 E/E_0$, and $s_1, s_2, \tau$ in units of $\gamma/a = 2\tau_{\rm GL}$,
\begin{equation}
    S_{\mu\nu}(\mathbf k,\omega\to0;E) = 32\,k_BT\,\sigma^{\rm AL} \int_0^\infty \dd s_1\dd s_2\dd\tau\; \frac{e^{\mathcal A + |\mathbf b|^2/8\Sigma}}{\Sigma}\, V_{\mu\nu}, \qquad \Sigma = s_1+s_2+\tau,
    \label{app:masternoneq}
\end{equation}
\begin{equation}
    \mathcal A = -2\Sigma(1+\kappa^2) + 2\kappa_\parallel f(s_1^2 - s_2^2) - \tfrac23 f^2(s_1^3+s_2^3+\tau^3), \qquad
    \mathbf b = \big(2fB - 4(s_1{-}s_2)\kappa_\parallel,\; -4(s_1{-}s_2)\kappa_\perp\big),
\end{equation}
with $B = s_1^2+s_2^2-\tau^2$ and the vertex moments $V_{xx} = \tfrac1{4\Sigma} + \mu_x^2 + f\tau\mu_x$, $V_{yy} = \tfrac1{4\Sigma} + \mu_y^2$, $\bm\mu = \mathbf b/4\Sigma$. The prefactor carries two internal checks: (i) at $E = 0$, $k = 0$ the triple integral evaluates elementarily to $\int_0^\infty \dd u\,(u^2/2)e^{-2u}/(4u^2) = 1/16$, so that $S_{\mu\nu} = 2k_BT\sigma^{\rm AL}\delta_{\mu\nu}$---the fluctuation-dissipation theorem emerges with the correct coefficient; (ii) at $E=0$ and finite $\kappa$ it reproduces $S_{T,L} = 2k_BT\sigma^{\rm AL}F_{T,L}(\kappa)$ with the closed-form scaling functions of Sec.~\ref{sec:nonlocal} (verified numerically to $10^{-4}$), while at $k=0$ the associated mean current reproduces Dorsey's $\Sigma_+$ exactly. The FDT ratios and anisotropies plotted in Figs.~\ref{fig:fdt} and \ref{fig:noneqk} follow by direct numerical evaluation of Eq.~\eqref{app:masternoneq}; for an isotropic film $\mathbf J = \sigma(|\mathbf E|)\mathbf E$, so the transverse differential conductivity equals $\sigma$ identically, which is why $X_\perp$ is normalized by $\sigma$.

\section{Fluctuation diamagnetism}
\label{app:dia}

\textit{Susceptibility---}In a perpendicular field $B$ the quadratic fluctuation modes occupy Landau levels with eigenvalues $a + \hbar\omega_c(n+\tfrac12)$, $\omega_c = eB/mc$, and areal degeneracy $B/\Phi_0$. Applying the Euler-Maclaurin expansion to the Gaussian free energy $F = T(B/\Phi_0)\sum_n \ln[a + \hbar\omega_c(n+\tfrac12)]$ gives the field-dependent part $\delta F = Te^2B^2/(48\pi mc^2a)$ and hence the areal susceptibility per layer $\chi_A = -\partial^2\delta F/\partial B^2 = -Te^2\xi^2/12\pi\hbar^2c^2$, which with $e = 2e_0$ is Eq.~\eqref{eq:chiA}. The same result follows as the $Q\to0$ limit of the static transverse response kernel: the reactive part of the thin-film kernel of Ref.~\cite{galaktionov1995}, $B_G(\kappa,0) = 1-\sqrt{1+\kappa^2}\,\mathrm{arcsinh}(\kappa)/\kappa$, yields $\chi(Q) = \chi_A G(Q\xi/2)$ with $G(x) = (3/x^2)[\sqrt{1+x^2}\,\mathrm{arcsinh}(x)/x - 1]$, which we have verified independently against the static GL current-current correlator. The full kernel obeys $B_G(\kappa,\varpi) = B_G(\kappa,0) + \tfrac{i\varpi}4 F_T(\kappa,\varpi)$: the fluctuation diamagnetism and the AL conductivity are the reactive and dissipative parts of the same nonlocal response.

\textit{No-double-counting---}For a 2D layer, the bound current of an out-of-plane magnetization density is purely transverse, $K_T(\bfq) = icqM_z(\bfq)$, giving Eq.~\eqref{eq:nodouble}: the spectral density of magnetization noise is the transverse current noise divided by $c^2q^2$. Note the complementary roles of the two static objects: the (positive) spontaneous noise $\langle|M_z(\bfq)|^2\rangle = T\,\Pi_T(\bfq)/c^2q^2$ is set by the paramagnetic current correlator $\Pi_T$, while the (negative, diamagnetic) response is the gauge-invariant difference $\chi(Q) = [\Pi_T(Q) - \Pi_T(0)]/c^2Q^2$; both coexist consistently with the FDT, which ties the noise to the dissipative part of the response at finite frequency.

\textit{Edge fields---}Near the edge of a uniformly magnetized film the moment density heals as $M(x) = M_\infty[1 - 3\int_1^\infty e^{-2 x u/\xi}\sqrt{u^2-1}\,u^{-4}\dd u]$ \cite{galaktionov1995}, where $M_\infty$ is the (fluctuational) diamagnetic moment in the bulk. The profile $M(x)$ rises linearly over $\xi(T)/2$; at heights above the sample $z \gg \xi/2$ the edge acts as a line current $I = cM_A$ and the estimates of Eq.~\eqref{eq:dBedge} follow, while for $z \lesssim \xi/2$ the profile smoothing (and hence $\xi(T)$ itself) becomes directly observable in a dc field map.

\end{widetext}
\end{document}